\documentclass[11pt]{article}
\usepackage{amssymb}
\usepackage{amsmath}
\usepackage{graphicx}

\def\beq{\begin{eqnarray}}
\def\eeq{\end{eqnarray}}
\def\ln{\,\mbox{ln}\,}

\def\sTr{\,\mbox{sTr}\,}

\def\Box{\square}

\def\al{\alpha}
\def\be{\beta}

\def\ga{\gamma}
\def\vp{\varepsilon}
\def\ep{\epsilon}

\def\la{\lambda}
\def\na{\nabla}

\def\rh{\rho}
\def\si{\sigma}

\def\ph{\varphi}
\def\ta{\tau}

\def\Ga{\Gamma}

\begin{document}


\begin{center}

{\large\sc On the Renormalization of CPT/Lorentz Violating
QED in Curved Space}
\vskip 3mm

{\small \bf Guilherme de Berredo-Peixoto$^{a,b}$} \footnote{E-mail
address: guilherme@fisica.ufjf.br},
\qquad
{\small \bf Ilya L. Shapiro$^{b}$}
\footnote{Also at Tomsk State Pedagogical
University, Russia. E-mail address: shapiro@fisica.ufjf.br}
\vskip 2mm

{\small\sl a. Departamento de Engenharia Rural, CAV,
Universidade do Estado de Santa Catarina,} \\
{\small\sl Lages, 88520-000,  SC, Brazil}

{\small\sl b. Departamento de F\'{\i}sica -- ICE,
Universidade Federal de Juiz de Fora,} \\
{\small\sl Juiz de Fora, 36036-330, MG, Brazil}

\end{center}
\vskip 5mm

\centerline{\uppercase{abstract}}
\begin{quotation}
\noindent
We consider the one-loop renormalization of QED in
curved space-time with additional Lorentz and/or CPT breaking
terms. The renormalization group equations in the vacuum sector
are derived. In the special case of Minkowski metric and with
constant Lorentz and CPT breaking terms these equations reduce to
the ones obtained earlier by other authors. The necessary form of
the vacuum counterterms indicate possible violations of the space
or time homogeneity or space isotropy in the gravitational
phenomena. However, the necessity of the phenomenologically most
interesting terms such as linear in the space-time curvature or
torsion, is related to the non-constant nature of the
dimensionless Lorentz and CPT breaking parameters.
\vskip 2mm

\noindent {\it Keywords: CPT and Lorentz violation, Curved space,
Renormalization.} \
\\
{\it PACS:} 04.62.+v,  12.90.+b.
\end{quotation}

\vskip 4mm                                                  %
\section{\large\bf  Introduction}                           %

One of the most perspective windows into the Planck-scale physics
is due to the possible violations of the fundamental (Lorentz
and/or CPT) symmetries \cite{LCPT1}. According to the existing
theoretical investigations \cite{LCPT2}, it might happen that, due
to the string scale quantum effects, the vacuum is not Lorentz
invariant. This means that in our part of the Universe there is
some fixed external field or fields which define a preferable
direction in the space-time. Similar vacuum effects may be
responsible for the
CPT violations. The conventional approach for the study of the
possible Lorentz and CPT violations is to consider the most
general form for these violations and then look for their
phenomenological consequences. The most promising type of
experiments here belong to the area of atomic physics
\cite{CPT-atom}, but there are also interesting options in the
area of high energy and solid state physics \cite{hep-CPT},
neutrinos experiments \cite{neutrinos},
gravitation \cite{Kost-grav} and cosmology \cite{CPTL-cosm}
(see further references therein).

From the quantum field theory viewpoint, the presence of
additional external fields means that the changes may occur not
only at the tree level but also in the loop corrections. The
investigation of the loop correction in CPT and Lorentz violated
QED has been started recently in the work \cite{Kost-loop}, where
the renormalization-group $\beta$-functions have been derived and
the corresponding quantum effects were indicated. One of the most
important aspects of quantum consideration is that it enables one
to elaborate the criteria of consistency for the theory. These
criteria concern the minimal necessary set of the terms in the
classical action which are usually required by renormalizability
and unitarity of the theory and also the magnitudes of the new
parameters in the actions, which should be consistent with the
renormalization group running caused by the loop effects. An
example of the theory where these criteria were very well
investigated is the quantum field theory in curved space-time. In
this case the consistency is achieved by introducing the so-called
vacuum action and the non-minimal interaction terms for the scalar
fields. The most remarkable fact is that the vacuum action must
include, along with the Einstein-Hilbert term, also the
cosmological constant and the full set of local fourth derivative
terms (see, e.g. \cite{book} for the introduction and further
references). This aspect of the theory leads to important
consequences for the gravitational phenomenology and, in
particular, is relevant for a better understanding of the
Cosmological Constant Problem \cite{nova}.

Indeed, one can consider the situation with the Lorentz and
CPT violating terms in QED (or any other quantum field theory)
from a similar viewpoint. Introducing new terms in the
matter sector one meets, at the quantum level, the necessity
to extend the classical action of vacuum by introducing
the full set of the new terms depending on Lorentz
and CPT violating parameters. The form of the necessary new
terms may be established on the basis of general considerations
using dimension and covariance. However, only practical one-loop
calculations may indicate which of those terms are the most
important ones. Let us remember that, due to the weakness of
the Lorentz and CPT violating terms, the higher loop vacuum
effects are definitely very small. Therefore the most important
terms are those which will emerge in the one-loop counterterms.
Hence the correct strategy in investigating the possible role
of Lorentz and CPT violating parameters in the gravitational
sector is to start by deriving the one-loop divergences for
the quantum field theory on curved background, including the
Lorentz and CPT violating terms in the matter section.
The results of such calculation can be used as a hint, indicating
the most important new terms to be implemented
into the gravitational action. At the next stage one has
to look for the possible phenomenological manifestation of these
new terms.

In this paper we report on the particular results of the
one-loop calculations in the Lorentz and CPT violating QED
in curved space-time. This calculation helps us to identify
the most interesting gravitational terms, that may be a good
starting point for the phenomenological analysis.

The paper is organized as follows. In section 2 we consider
the Lorentz and CPT violating QED in curved space-time and
derive the most relevant one-loop counterterms. Our results
are less general than the ones of the previous calculation
in flat space-time \cite{Kost-loop}, because we set some of
the new parameters to zero. The inclusion of these parameters
leads to enormous increase of the size of the calculational
work and we postpone this calculation for the future. On
the other hand, we perform a more general calculation that
the one of \cite{Kost-loop}, because we do not request nor
that the Lorentz and CPT violating parameters are
constants\footnote{The possibility of the space-time
dependent Lorentz-violating parameters has been considered
in \cite{MPV}. },
neither the flatness of the space-time.
In this way we achieve, in section 3, the form of the
qualitatively new terms in the gravitational action
and consider the renormalization group for (some of) the
corresponding parameters. The
section 4 includes a general discussion of their possible
physical effects and in section 5 we draw our conclusions.

\section{\large\bf  Derivation of the one-loop divergences}

Our starting point is the action describing an
extended version of QED with Lorentz and CPT symmetry breaking
terms, minimally coupled to gravitation
\beq
S & = & \int d^4x\sqrt{-g}\,\left\{\,
\frac{i}{2}\,\bar{\psi}\Ga^\mu D_\mu\psi
- \frac{i}{2}\, D^{\star}_\mu\bar{\psi}\Ga^\mu\psi
-\bar{\psi}\, M\,\psi -\frac{1}{4} F_{\mu\nu}\,F^{\mu\nu}
\right.
\nonumber
\\
& - & \left.
\frac{1}{4}\, (k_F)_{\mu\nu\al\be}\,F^{\mu\nu}\,F^{\al\be}
+ \frac{1}{2}\,(k_{AF})^\rh \,\vp_{\rh\la\mu\nu}
\,A^\la\,F^{\mu\nu}\,\right\}\,.
\label{action}
\eeq
Here
\beq
D_\mu = \na_\mu + i\,q\,A_\mu\, ; \;\;\; D^{\star}_\mu
= \na_\mu - i\,q\,A_\mu\, ; \;\;\;
\Ga^\nu = \ga^\nu + \Ga_1^\nu\, ; \;\;\; M = m + M_1\, ,
\eeq
the operator $\na_\mu$ is the covariant derivative,
$F_{\mu\nu} = \na_\mu A_\nu - \na_\nu A_\mu$ and the
quantities $\Ga_1^\nu$ and $M_1$ are given by
\beq
\Ga_1^\nu & = & c^{\mu\nu}\ga_\mu + d^{\mu\nu} \ga_5\ga_\mu
+ e^\nu + i\,f^\nu\ga_5
+\frac{1}{2}\,g^{\la\mu\nu}\si_{\la\mu}\, , \\
M_1 & = & a_\mu\,\ga^\mu + b_\mu\,\ga_5\,\ga^\mu
+ i\, m_5\ga_5 + \frac{1}{2}\, H_{\mu\nu}\,\si^{\mu\nu}\, .
\label{break}
\eeq
Along with the parameters $(k_F)_{\mu\nu\al\be}$ and $(k_{AF})^\rh$,
the quantities
$a_\mu$, $b_\mu$, $m_5$, $c^{\mu\nu}$, $d^{\mu\nu}$, $e^\mu$,
$f^\mu$,
$g^{\la\mu\nu}$ and $H_{\mu\nu}$ are CPT and/or Lorentz violating
parameters.
An extensive discussion of the possible origin of these parameters
and also their numerous phenomenological implications can be found
in \cite{LCPT1,LCPT2} and we will not consider these aspects
here. Let us only notice that the quantities constituting $\Ga_1^\nu$
and the parameter $(k_F)_{\mu\nu\al\be}$ are dimensionless and all
other parameters have non-zero dimensions which can be easily
found from eq. (\ref{action}).

Our purpose is to investigate the one-loop renormalization in
the vacuum sector, depending on the metric and on the set of the
CPT and/or Lorentz violating parameters. As usual, we assume that
these new fields are small, therefore our calculation can be
restricted to the first order in the Lorentz and CPT violating
parameters. At the same time, we do not see why these parameters
must be constants and hence we will let them being local functions
of the space-time coordinates. Indeed, the space and time
derivatives of these parameters should be small, such that the
linear approximation in the new parameters and their derivatives
remain sufficient.

In order to investigate the one-loop renormalization of the
Lorentz/CPT breaking coefficients, we consider the background
field method splitting,
\beq
\psi \to \psi + \eta \, ; \;\;\;\; \bar{\psi}\to\bar{\psi}
+ \bar{\eta}\, ;\;\;\;\; A_\mu\to A_\mu + B_\mu\, ,
\eeq
where $\eta$, $\bar{\eta}$ and $B_\mu$ are the quantum fields,
and calculate the complete one-loop divergences of theory
(\ref{action}), given by\footnote{The definition of the
operation of supertrace $\,\sTr\,$ is given below,
after the eq. (\ref{h}).}
\beq
\Ga^{(1)}_{{\rm div}}
= \frac{i}{2} \sTr\ln \hat{\cal{H}}|_{{\rm div}}\,.
\eeq
where the operator $\hat{\cal{H}}$ is the operator of the
bilinear in quantum fields part of the action together with
the gauge-fixing term. Let us introduce this term in the form
\beq
S_{gf}=\frac{1}{2\al}\,\int d^4x\sqrt{-g}\,
\left(\na_\mu B^\mu\right)^2\,,
\label{gauge fixing}
\eeq
where $\al$ is an arbitrary parameter of the gauge fixing.
Let us notice that, with this choice of the gauge-fixing,
the corresponding Faddeev-Popov ghosts contribute only to
the vacuum (metric-dependent) sector of the theory and
these contributions do not depend on the new Lorentz
breaking parameters of the theory. Therefore one can
safely disregard the contribution of the ghosts except
when we consider these terms in eq. (\ref{vacuum}).
The use of the background field method (see \cite{book}
for the introduction and further references) guarantees
the gauge invariance of the quantum corrections.

For practical calculations we choose the value of \ $\al=-1$,
such that without the Lorentz breaking parameters, the
bilinear form of the action has the most simple
minimal form, with second derivatives forming the
$\,\Box=g^{\mu\nu}\na_\mu\na_\nu\,$ combination
(see, for example, the expression (\ref{operator}) below,
in the case  $\,\hat{\cal{K}}^{\rh\la}=0 $).
The natural question is whether the quantum corrections
may depend on the choice of parameter $\al$. As far as we
are interested in the one-loop divergences and related
objects like \ $\be$-functions, the well known general
theorems \cite{LTV,WeinbergQFT} tell us that the gauge
fixing dependence should disappear on the classical
mass-shell. As a result, only the renormalization of the
wave function for the fermion may depend on the choice
of \ $\al$, while the \ $\be$-functions for all parameters
(including the Lorentz-violating ones) are not affected
by this dependence.

After performing the bilinear expansion, it proves useful
to make a change of variables
$$
\eta\,=\,-\,\frac{i}{2}\,\left(\ga^\mu\na_\mu-im\right)\,\chi\,.
$$
The corresponding Jacobian depends only on the metric and can
be evaluated using the known algorithms (see, e.g. \cite{book}).
After the mentioned change of quantum variables the bilinear
part of the total action acquires the form
\beq
S = \frac{1}{2}\int d^4x\sqrt{-g}\left(
\begin{array}{ccc} B_\mu & | & \bar{\eta} \end{array} \right)
\,\hat{\cal{H}}\,\left( \begin{array}{c}
B_\nu \\ \chi \end{array} \right)\, .
\eeq
Let us notice that the above expression has an unusual form,
making practical calculations rather difficult. Due to the
presence of the massless parameters in $\Ga_1^\nu$ and also
$(k_F)_{\mu\nu\al\be}$ in the action (\ref{action}), the
differential operator $\hat{\cal{H}}$ has a general non-minimal
structure
\beq
\hat{\cal{H}} = \hat{\cal{K}}^{\rh\la}\,\na_\rh \na_\la
+ \hat{\mathbf 1}\,\Box
+ 2\,(\hat{h}_0^\rh + \hat{L}^\rh )\,\na_\rh + (\hat{\Pi}_0
+ \hat{\cal{M}})\, ,
\label{operator}
\eeq
where we have separated the conventional terms and the new,
(unusual) ones. In particular,
$$
\hat{\mathbf 1} = \left(\begin{array}{ccc}
g^{\mu\nu} & | & 0 \\ 0 & | & I \end{array}\right)\,,
$$
with $I$ being the identity operator in the fermionic sector,
$\hat{h}_0^\rh$ and $\hat{\Pi}_0$ are those parts of the
corresponding expressions (matrices), which are independent
on the Lorentz violating coefficients. These formulas simply
correspond to the pure QED theory in curved space-time.
Their explicit form is
\beq
\hat{h}_0^\rh = \left( \begin{array}{ccc} 0 &
\quad
\frac{i}{2} q\bar{\psi}\ga^\mu\ga^\rh
\\
0 &
\quad
\frac{i}{2} q\ga^\al\ga^\rh A_\al
\end{array} \right)
\,, \qquad
\hat{\Pi}_0 = \left( \begin{array}{ccc} -R^{\mu\nu} &
\quad
m\,q\,\bar{\psi}\,\ga^\mu
\\
-2\,q\,\ga^\nu\,\psi &
\quad
m^2I -\frac{I}{4}\,R + m\,q\,\ga^\al\,A_\al
\end{array} \right)\,.
\eeq
Furthermore, the matrices $\hat{\cal{K}}^{\rh\la}$,
$\hat{L}^\rh$ and $\hat{\cal{M}}$ are linear on the
Lorentz violating coefficients,
\beq
\hat{\cal{K}}^{\rh\la} = \left(\begin{array}{ccc}
-2\,k_F^{\mu (\rh\la )\nu}  &  \quad  0
\\
0 & \quad  \Ga_1^{(\rh}\,\ga^{\la )}
\end{array}\right)\,,
\eeq
\beq
\hat{L}^\rh = \left(\begin{array}{ccc}
\na_\ta k_F^{\ta\mu\rh\nu} - (k_{AF})_\ta\,\vp^{\ta\mu\nu\rh} &
\quad
\frac{i}{2}\,q\,\bar{\psi}\,\Ga_1^\mu\,\ga^\rh \\
0 & \quad
\frac{i}{2}\,q\,\Ga_1^\al\,\ga^\rh\,A_\al -
\frac{i}{2}\,m\,\Ga_1^\rh + \frac{i}{2}\,M_1\,\ga^\rh +
\frac{1}{4}\,\na_\mu\Ga_1^\mu\,\ga^\rh
\end{array}\right)
\eeq
and
\beq
\hat{\cal{M}} = \left(\begin{array}{ccc}
\na_\rho k_{AF}^\la \vp_\la\mbox{}^{\mu\rho\nu}
-k_F^{\rh\mu\la\ta}R^{\nu}\mbox{}_{\ta\rh\la}
&  \,\,\,
mq\bar{\psi}\Ga_1^\mu
\\
-2\,q\Ga_1^\nu\psi
&  \,\,\,
m \big(
q\Ga_1^\rh A_\rh + M_1 -\frac{i}{2}\na_\al\Ga_1^\al
\big)
+ \frac{1}{2}\Ga_1^\al\ga^\be [\na_\al,\na_\be]
\end{array}\right).
\label{M}
\eeq
As we have already mentioned above,
the operator (\ref{operator}) has a non-minimal form and
therefore the standard Schwinger-DeWitt technique for
deriving the divergences can not be applied. In this
situation one can either use local momentum representation
(see, e.g. \cite{parker-toms}) or apply the
generalized Schwinger-DeWitt method \cite{bavi},
making expansion into
the new Lorentz/CPT breaking terms. The last calculation
is definitely possible, in particular we could obtain the
general expression for the divergences of the theory.
Unfortunately, from the technical point of view, it is rather
difficult, demanding extremely long and complicated
algebraic computer calculations. At the
same time, the most important physical information can be indeed
obtained in a relatively economic way. Therefore we postpone
the complete calculus for the possible next work and concentrate
here on the reduced version. One can simplify the problem by
setting $\hat{\cal{K}}^{\rh\la} =0$, reducing the
quadratic operator to the minimal form. This restriction can
be achieved by setting
$k_F^{\mu (\rh\la )\nu} = 0$ and $\Ga_1^{\rh} = 0$. In fact,
$k_F^{\mu (\rh\la )\nu}$ shows up also in the expression for
$\hat{L}^\rh$. In order to have an extra qualitative information,
in what follows we will not simply set
$k_F^{\mu (\rh\la )\nu} = 0$,
but just remember that the corresponding terms can be modified
by taking into account the
$\hat{\cal{K}}^{\rh\la}$-dependent terms.

In the framework of the Schwinger-DeWitt technique, we find the
one-loop divergences by using the known formula (see, e.g.
\cite{book})
\beq
\Ga^{(1)}_{{\rm div}} =-
\frac{\mu^\ep}{(4\pi)^2\,\ep}\,\int\,d^nx\sqrt{-g}
\,\,{\rm sTr}\,\left\{\, \frac{1}{2}\hat{P}^2
+ \frac{1}{12}\hat{S}_{\al\be}\hat{S}^{\al\be}
+ \mbox{vacuum terms}\,\right\}\,,
\label{general}
\eeq
where $\ep = n-4$ is the dimensional regularization parameter and
\beq
\hat{P} & = & \hat{\Pi} + \frac{1}{6}\,\hat{\mathbf 1}\,R
- \na_\mu \hat{h}^\mu - \hat{h}_\mu\hat{h}^\mu\,,
\label{P}
\eeq
\beq
\hat{S}_{\al\be} & = & [\na_\be\,,\,\na_\al ] \hat{\mathbf 1}
+ \na_\be\hat{h}_\al - \na_\al\hat{h}_\be
+ \hat{h}_\be\hat{h}_\al - \hat{h}_\al\hat{h}_\be\,,
\label{S}
\eeq
with
\beq
\hat{h}^\rh = \hat{h}_0^\rh + \hat{L}^\rh\,;\;\;\;\;\;\;\;
\hat{\Pi} = \hat{\Pi}_0 + \hat{\cal{M}}\, .
\label{h}
\eeq
Finally, the symbol \ ${\rm sTr}$ \ in the formula
(\ref{general}) stands for the supertrace
(or Berezinian). This means that the trace of the matrix
operator in (\ref{general}) and in what follows
is an algebraic sum of the traces of the diagonal matrix
elements, with the signs taken according to the statistics
of the corresponding quantum fields. In particular, in the
bosonic sector of the fields \ $B^\mu$ \ the coefficient
of the contribution is \ $+1$ \ and in the fermionic
\ $\bar{\eta}\chi$ - sector the coefficient is \ $-2$.

In the expression (\ref{general}) we have disregarded those
terms which depend only on the metric. The reason is that they
are given by the sums of the contributions of the free fields
which are very well-known (see, e.g. \cite{birdav,book}).
After performing some algebra, we obtain, in up to the
first order in the new parameters,
\beq
\frac{1}{2}\, \sTr\,\hat{P}^2 & = &
\frac{1}{2}\, \sTr\,\hat{P}_0^2
+ \sTr\,\left( \hat{P}_0\,\hat{\cal{M}}
- \hat{P}_0\,\na_\nu\hat{L}^\nu
- \hat{P}_0\,\hat{h}_0^\nu\,\hat{L}_\nu
- \hat{P}_0\,\hat{L}_\nu\,\hat{h}_0^\nu
\right) \, ,
\nonumber
\\
\sTr\,\hat{S}_{\al\be}\hat{S}^{\al\be}
& = &
\sTr\,\left(\,\hat{S}_{0\,\al\be}\hat{S}_0^{\al\be}
+ 4\,\hat{S}_{0\,\al\be}\,\na^\be\,\hat{L}^\al
+ 4\,\hat{S}_{0\,\al\be}\,
[\hat{h}_0^\be\,,\,\hat{L}^\al ]\,\right)\,,
\nonumber
\eeq
where the operators $\hat{P}_0$ and $\hat{S}_{0\,\al\be}$
correspond to the pure QED. These operators can be obtained
from the expressions (\ref{P}) and (\ref{S}) by setting
all Lorentz/CPT breaking parameters to zero.
The final result for the divergent part of the one-loop
effective action has the form
\beq
\Ga^{(1)}_{div} & = &
-\frac{\mu^\ep}{(4\pi)^2\,\ep}\,\int\,d^nx\,\sqrt{-g}\,
\left\{\,- \,2\,m\,q\,H^{\mu\nu}F_{\mu\nu}
- \frac{4}{3}\,q\,(\na_\mu a_\nu)\,F^{\mu\nu}
\right.
\nonumber \\
& - & \left.
2q^2\,\bar{\psi}\,\Big[\, a_\mu\ga^\mu
+ \Big(b_\mu + \frac12\,S_\mu -3(k_{AF})_\mu\Big)\ga_5\ga^\mu
+ 4im_5\ga_5 + \frac{i}{2}\na_\ta
(k_F)^\ta\mbox{}_\mu\ga^\mu
\Big]\,\psi
\right.
\nonumber \\
& + & \left.
R_{\mu\al}\na_\rh\na_\ta k_F^{\ta\al\rh\mu}
-\frac{1}{6}R\na_\mu\na_\nu k_F^{\mu\nu}
+ \frac{1}{3}R_{\mu\rho\al\be}\na^\be\na_\ta k_F^{\ta\mu\al\rho}
\right.
\nonumber \\
& - & \left.
\frac{1}{12}k_F^{\rho\la\mu\ta}RR_{\rho\la\mu\ta}
+ \frac12 k_F^{\al\rho\ta\la}R^\mu\mbox{}_{\rho\ta\la}R_{\mu\al}
\,\right\} + \Ga^{(1)}_{QED\; div}
+ \Ga^{(1)}_{vac}\left[g_{\mu\nu}\right]\,,
\label{div}
\eeq
where we used notations
\ \ $(k_F)^{\mu\la\nu}\mbox{}_\la = (k_F)^{\mu\nu}$,
\ \ $S_\mu = \na_\al k_F^{\al\rho\la\si}\vp_{\mu\rho\la\si}$
\ \ and
\beq
\Ga^{(1)}_{QED\; div} =
-\frac{\mu^\ep}{(4\pi)^2\,\ep}\,\int\,d^nx\,\sqrt{-g}\,
\left\{\, -\frac{1}{3}\,q^2\,F_{\mu\nu}F^{\mu\nu} +
2\,i\,q^2\,\bar{\psi}\,(\ga^\mu\,D_\mu + 4\,i\,m)\,\psi \right\}
\label{QED div}
\eeq
is the one-loop divergences for the pure QED theory. Finally,
\ $\Ga^{(1)}_{vac}\left[g_{\mu\nu}\right]$ \ is the divergent
part of the metric-dependent vacuum effective action of QED
(taking the ghosts contributions into account)
\beq
\Ga^{(1)}_{vac}\left[g_{\mu\nu}\right] &=&
-\frac{\mu^\ep}{(4\pi)^2\,\ep}\,\int\,d^nx\,\sqrt{-g}\,
\left\{\,\frac{3}{20}\,C^2_{\mu\nu\al\be}
- \frac{73}{360}\,E + \frac{2}{3}\,m^2 - 4\,m^4 \,\right\}\,,
\label{vacuum}
\eeq
with $C^2_{\mu\nu\al\be}$ and $E$ representing the square of the
Weyl tensor and the Gauss-Bonnet topological term (Euler
characteristic). In the expressions presented
above we did not include total derivatives. It is
easy to see that the above expression includes, along with
the usual QED divergences, many new terms. Part of these terms,
concentrated in the last two lines of eq. (\ref{div})
can be characterized as the vacuum ones, because they do not
depend on the matter fields. Other new terms have a non-minimal
form, analogous to the $R\ph^2$-term for the scalar field theory
in curved space. Let us notice that there is no divergent
Chern-Simons - like term, $b_\mu A_\nu F_{\al\be}\vp^{\mu\nu\al\be}$.
This term is possible due to the dimensional reasons, however it
does not show up in the one-loop counterterms.

\section{General discussion of the one-loop renormalization}

Let us start the analysis of the result (\ref{div}) by making
comparison with the available particular results. First of all,
eq. (\ref{QED div}) has the standard well-known form of the QED
divergences in curved space-time. This is perfectly consistent
with the general features of the renormalization in curved
space-time (see, e.g. \cite{book}). The renormalization of
the minimal sector performs independent on external fields.
We can conclude
that this feature holds in the present case of a Lorentz/CPT
violating theory.

As a second step, consider the particular case where
the background metric is the flat Minkowski one and moreover
assume that all Lorentz violation parameters
are constants. This is exactly the case investigated in
\cite{Kost-loop}. Unfortunately we are not able to make a
full comparison, because \cite{Kost-loop} treats the general
case with an arbitrary matrix $\hat{\cal{K}}^{\rh\la}$
\footnote{At the same time, the result of \cite{Kost-loop}
is restricted to the flat space, while we are working in
the general curved one and also consider the non-constant
Lorentz/CPT violation parameters.}. However, even the
partial comparison of the two calculations is useful and
important. The one-loop divergences in this case reduce to
\beq
\Ga^{(1)}_{div} &=& \Ga^{(1)}_{QED\; div}
\nonumber
\\
&+& \frac{2q^2\,\mu^\ep}{(4\pi)^2\ep}\,
\int d^nx \sqrt{-g}\,\left\{ \bar{\psi}\,\Big[
a_\mu\ga^\mu + b_\mu\ga_5\ga^\mu
+ 4im_5\ga_5 - 3(k_{AF})_\mu\ga_5\ga^\mu \Big]\,\psi\right\}\,.
\label{flat}
\eeq
Let us notice that the term
\ $- \,2\,m\,q\,H^{\mu\nu}F_{\mu\nu}$ \ is not included here
because for $\,H^{\mu\nu}=const\,$ it is a total derivative.
The expression (\ref{flat}) corresponds to the one which has
been used
in \cite{Kost-loop} for deriving the renormalization group
equations. Correspondingly, the proper equations (see below)
reduce to the ones of \cite{Kost-loop} in the corresponding
limit.

Let us now consider the qualitatively new terms due to the
nontrivial metric, which have no analogs in the flat space
case. The form of divergences in the fermionic sector shows
the usual renormalization of the electromagnetic-like
and torsion-like terms, with the $a^\mu$ and $b^\mu$
coefficients correspondingly. The form of the counterterms
is controlled by the gauge symmetries, as it was discussed
in \cite{betor,guhesh,torsi}. Furthermore, the axial vector
torsion component $b^\mu$ gains new contributions from the
new $S^\mu$ and $k_{AF}^\mu$ fields. It is remarkable that
the $S^\mu$ contribution is caused by the derivative of the
massless $k_{F}^{\mu\nu\al\be}$ field. Therefore if this
field is present and is not  exactly constant, the torsion
$b^\mu$ term should be indeed present in the classical
action, otherwise the theory is not renormalizable. Similar
considerations apply to the torsion-like trace vector
$a^\mu$. However, here we meet an interesting special
situation, because the contribution from
\ $\na_\tau (k_F)^\tau_\mu$ \ has an extra imaginary unit.
In fact, this form of the
imaginary term exactly corresponds to the torsion trace
\cite{torsi} interacting to fermion, but in the torsion
case this term always cancels. Perhaps, the same
concerns the corresponding quantum contributions in
(\ref{div}), because they may be actually cancelled by
another \ $\na_\tau (k_F)^\tau_\mu$-terms coming from
the \ $\cal{O}(\hat{\cal{K}}^{\rh\la})$ \
contributions which we did not calculate here.

In the curvature-dependent
sector the situation is also quite interesting.
On dimensional reasons one could expect the counterterms of the
form
\beq
S_{{\rm full}} = \int d^4x\sqrt{-g}\,
\left\{\phi R + \phi^{\mu\nu} R_{\mu\nu}
+ \phi^{\mu\nu\al\be} R_{\mu\nu\al\be}\right\}
+ S_{GHD}\,,
\label{linear curvature}
\eeq
where the last part $\,S_{GHD}\,$ represents generalized
form of the higher derivative term. Let us remember that the
renormalizable theory in curved space-time always includes
higher derivative vacuum (metric dependent) terms (see, e.g.
\cite{birdav,book}. In case when some extra field (e.g.
torsion) are present, the consistent form of the vacuum
action becomes more complicated, involving the dependence of
these extra fields. For example, in case of torsion the total
number of possible independent structures in the vacuum action
equals to 168 \cite{chris}. It is obvious that the number
of necessary structures in the present case, with all new
(dimensional and dimensionless) fields will be enormous
and therefore we will not try to list them. Instead we
just notice that, in our one-loop calculations, only two
such terms emerge as counterterms and therefore
\beq
S_{GHD}\,=\, \int d^4x\sqrt{-g}\,
\left\{\eta_1^{\mu\nu\al\be}\,R\,R_{\mu\nu\al\be} +
\eta_2^{\mu\nu\al\be}\,R_\mu^\la\,R_{\la\nu\al\be}\right\}
\,+\,...\,\,.
\label{square curvature}
\eeq
In the expressions (\ref{linear curvature}) and
(\ref{square curvature}), the coefficients
$\,\phi ,\,\phi^{\mu\nu} ,\,\phi^{\mu\nu\al\be},\,
\eta_1^{\mu\nu\al\be},\,\eta_2^{\mu\nu\al\be}\,$ should
depend on the new Lorentz/CPT breaking parameters in (\ref{break}).
The renormalizability of the theory requires that
these terms should be introduced into the classical gravitational
action, along with the usual Einstein-Hilbert, cosmological and
higher derivative terms. Let us notice that the first term in
(\ref{linear curvature}) has the known metric-scalar
form\footnote{Taken alone, this action is conformally equivalent
to the Brans-Dicke action and to many other versions of the
metric-scalar theory. With other new terms present, this
equivalence does not hold.}, while the next two terms have a
qualitatively new structure and
may lead to new physical effects. Recently, an extensive
investigation of the possible phenomenological manifestations
of the new terms (\ref{linear curvature}) has been performed
in \cite{Kost-post-New} on the basis of the post-Newtonian
approximation. According to this work, the
phenomenological manifestations of these, linear in
curvature, terms are potentially measurable and therefore the
theoretical status of these terms and especially their relation
to the possible violation of the Lorentz/CPT symmetries in the
matter sector deserve our attention. The form of divergences
(\ref{div}) indicates that the linear in curvatures
terms do renormalize only due to the
non-constant nature of the new parameters in the matter
sector. Let us remark that this result does not depend on the
assumption  $\hat{\cal{K}}^{\rh\la}=0$ which we have used in
the present work, because the $\hat{\cal{K}}^{\rh\la}$
matrix is composed by the dimensionless parameters and
therefore can contribute to the dimensional coefficients
$\,\phi ,\,\phi^{\mu\nu} ,\,\phi^{\mu\nu\al\be}\,$ only
through the derivatives of these parameters. All in all,
the possible sources of the $\,\phi ,\,\phi^{\mu\nu}\,$ and
$\,\phi^{\mu\nu\al\be}\,$ terms may be either the derivatives
of the dimensionless parameters in the matter sector or the
higher loop corrections. In the last case the numerical
values of the contributions would be extremely weak.

\section{Renormalization group}

The renormalization group equations in the matter sector have
the form which follow from the expression for the divergences
\beq
\mu\frac{da_\nu}{d\mu} = \frac{da_\nu}{dt}
 =  -\frac{i q^2}{16\pi^2}\,\na_\al (k_F)^\al\mbox{}_\nu\, ,
\label{a}
\\
\frac{db_\mu}{dt} = \frac{3q^2}{8\pi^2}\,(k_{AF})_\mu
- \frac{q^2}{16\pi^2}\, S_\mu\,,
\qquad
\frac{dH_{\mu\nu}}{dt} = \frac{q^2}{8\pi^2}\,H_{\mu\nu}\,,
\nonumber
\\
\frac{dk_{AF\mu}}{dt} = \frac{q^2}{6\pi^2}\, (k_{AF})_\mu\, ,
\qquad\qquad
\frac{dk_F^{\mu\nu\al\be}}{dt}
 = \frac{q^2}{6\pi^2}\,k_F^{\mu\nu\al\be}\, ,
\nonumber
\\
\frac{dm_5}{dt} = -\frac{3q^2}{8\pi^2}\,m_5\, .
\eeq
In the particular case when all parameters are constants,
these equations are exactly the ones obtained in \cite{Kost-loop}.
According to the general analysis \cite{book} these renormalizations
group equations preserve their form in the curved space-time as well.
The only difference is the interpretation of the renormalization
group parameter $t$. The definition given in eq. (\ref{a})
is
the general one for the $\overline{\rm MS}$ renormalization scheme.
At the same time, the standard physical interpretation in
the flat space is that $t$ is the energy-momentum scaling parameter,
while in curved space it is a metric scaling parameter \cite{book}
(see also further references therein). As it was already
indicated in \cite{Kost-loop}, the running of the parameters
is small and therefore it has no much sense to analyze the
renormalization group flows for these small parameters. Rather
than that, the equations presented above may be used as a hint
for seeing the
relation between different parameters, especially for the
case of non-constant ones. For instance, we observe that the
vector and axial vector parts of the torsion tensor $a^\mu$
and $b^\mu$ may be caused by the variable Lorentz violating $k_F$
field in the vector sector.

The renormalization group equations of the new interaction
parameters in the gravitational sector
$\phi_F^{\mu\nu}$, $\phi$, $\phi^{\mu\nu}$ and $\phi^{\mu\nu\al\be}$,
can be written as
\beq
\frac{d\phi_F^{\mu\nu}}{dt}
 =  \frac{q^2}{12\pi^2}\,\phi_F^{\mu\nu}
+ \frac{mq}{8\pi^2}\,H^{\mu\nu} + \frac{q}{12\pi^2}\,\na^\mu a^\nu\,,
\nonumber
\eeq
\beq
\frac{d\phi}{dt} = \frac{1}{96\pi^2}\,\na_\mu\na_\nu k_F^{\mu\nu}\,,
\qquad
\frac{d\phi^{\mu\nu}}{dt} =
-\frac{1}{16\pi^2}\, \na_\al \na_\be k_F^{\al\mu\be\nu}\, ,
\qquad
\frac{d\phi^{\mu\nu\al\be}}{dt}
=  -\frac{1}{48\pi^2}\,\na^\be\na_\la k_F^{\la\mu\al\nu}\,.
\nonumber
\eeq
\beq
\frac{d\eta_1^{\mu\nu\al\be}}{dt}
 =  \frac{1}{192\pi^2}\, k_F^{\mu\nu\al\be}\,,
\qquad
\frac{d\eta_2^{\mu\nu\al\be}}{dt}
=  -\frac{1}{32\pi^2}\, k_F^{\mu\nu\al\be}\,.
\eeq
Here we can observe also the non-trivial nature of the variable
parameters in the matter fields sector. Perhaps the most interesting
is that only these variations are linked with the Lorentz violations
in the gravitational sector.

\section{Conclusions}

We have investigated the one loop renormalization of the
Lorentz/CPT violating QED in curved space-time, treating the
Lorentz/CPT violating parameters as fields rather than as
constants. The practical calculation has been performed for
the "minimal" sector of the theory, admitting application
of the usual Schwinger-DeWitt technique. We have found the
relation between the Lorentz violating terms in the matter
and gravitational sector. In particular, the corresponding
gravitational terms which were recently discussed in
\cite{Kost-post-New}, are necessary at the quantum level in
case when the Lorentz violating terms in the matter section
are not exactly constants.
\vskip 4mm


\noindent
{\large\bf Acknowledgments.} \

The authors are grateful to A. Kostelecky for stimulating
discussion
and correspondence. G.B.P. has been supported by the post-doctoral
fellowship from PRODOC/CAPES (Brazil). I.Sh. has been partially
supported by the research fellowships from CNPq (Brazil) and by
research grants from CNPq, FAPEMIG (Minas Gerais, Brazil) and
ICTP (Italy).




\begin {thebibliography}{99}

\bibitem{LCPT1}
V. A. Kostelecky, S. Samuel,
Phys. Rev. D39 (1989) 683;

A. Kostelecky, R. Potting,
Phys. Rev. D63 (2001) 046007.

\bibitem{LCPT2}
R. Jackiw, A. Kostelecky, Phys. Rev. Lett. 82 (1999) 3572;

D. Colladay, A. Kostelecky,
Phys.Rev. D55 (1997) 6760;
Phys.Rev. D58 (1998) 116002;

A. Kostelecky, R. Lehnert,
Phys. Rev. D63 (2001) 065008;

A. A. Andrianov, P. Giacconi, R. Soldati,
JHEP 0202 (2002) 030; hep-th/0110279.

J. Alfaro, A. A. Andrianov, M. Cambiaso, P. Giacconi
and R. Soldati, Phys. Lett. B639 (2006) 586, [hep-th/0604164].

\bibitem{CPT-atom}
For the recent review and further references see

A. Kostelecky, {\sl The search for relativity violations},
Sci. Am. 291 (2004) 75;

A. Kostelecky,
Presented at the Third Meeting on CPT and Lorentz Symmetry,
Bloomington,
Indiana, August 2004, [hep-ph/0412406];

R. Bluhm, {\sl Overview Of The Sme: Implications And
Phenomenology Of Lorentz Violation},
Talk at 339th WE Heraeus Seminar on Special Relativity, Potsdam,
Germany, 13-18 Feb 2005, [hep-ph/0506054];

D. Mattingly, {\sl Modern Tests Of Lorentz Invariance},
Living Rev. Rel. 8 (2005) 5, [gr-qc/0502097].

\bibitem{hep-CPT}
R. Bluhm, A. Kostelecky,
Phys. Rev. Lett. 84 (2000) 1381;

R. Bluhm, A. Kostelecky, C. Lane,
Phys. Rev. Lett. 84 (2000) 1098.

A. Kostelecky, C. Lane,
Phys. Rev. D60 (1999) 116010.

R. Bluhm, A. Kostelecky, N. Russell,
Phys. Rev. Lett. 82 (1999) 2254.

\bibitem{neutrinos}
H. Murayama and T. Yamagida, Phys. Lett. B520 (2001) 263-268;

G. Barenboim, L. Borissov, J.D. Lykken and A. Yu Smirnov,
JHEP 0210 (2002) 001;

A. Kostelecky and M. Mewes, Phys. Rev. D 70 (2004) 031902;
D69 (2004) 016005.

\bibitem{Kost-grav}
V. A. Kostelecky, Phys.Rev. D69 (2004) 105009, [hep-th/0312310].

R. Bluhm, A. Kostelecky,
Phys. Rev. D71 (2005) 065008, [hep-th/0412320].

\bibitem{CPTL-cosm}
O. Bertolami, D. Colladay, A. Kostelecký, R. Potting,
Phys. Lett. B395 (1997) 178.

A. Kostelecky, M. Mewes,
Phys. Rev. Lett. 87 (2001) 251304;
Phys. Rev. D66 (2002) 056005.

O. Bertolami, J. Paramos, S. G. Turyshev
359th WE-Heraeus Seminar on "Lasers, Clocks, and Drag-Free:
Technologies for Future Exploration in Space and Tests of Gravity,"
ZARM,
Bremen, Germany,
May-June, 2005; [gr-qc/0602016].

Ralf Lehnert,
Cosmology and spacetime symmetries,
(New Worlds in Astroparticle Physics, Faro, Portugal, January, 2005),
[hep-ph/0508316].

R. Bluhm, A. Kostelecky,
Phys. Rev. D71 (2005) 065008.

\bibitem{Kost-loop}
A. Kostelecky, C. Lane, A. Pickering,
Phys.Rev. D65 (2002) 056006, [hep-th/0111123];

V.Ch. Zhukovsky, A.E. Lobanov, E.M. Murchikova,
Phys. Rev. D73 (2006) 065016.

\bibitem{MPV} Manuel Perez-Victoria
JHEP 0104 (2001) 032, hep-th/0102021.

\bibitem{book} I.L. Buchbinder, S.D. Odintsov, I.L. Shapiro,
{\sl Effective Action in Quantum Gravity} (IOP Publishing,
Bristol, 1992).

\bibitem{nova} I.L. Shapiro, J. Sol\`{a},
JHEP 02 (2002) 006.

\bibitem{LTV}
B.L. Voronov, P.M. Lavrov and I.V. Tyutin,
Sov.J.Nucl.Phys. {\bf 36} (1982) 498;

J. Gomis and S. Weinberg, Nucl.Phys. {\bf B469} (1996) 473.

\bibitem{WeinbergQFT}
S. Weinberg, {\sl The Quantum Theory of Fields:
Foundations.} (Cambridge Univ. Press, 1995).

\bibitem{parker-toms} L. Parker and D.J. Toms,
Phys. Rev. D29 (1984) 1584.

\bibitem{bavi}  A.O. Barvinsky and G.A. Vilkovisky,
The generalized Schwinger-DeWitt technique in gauge theories and
quantum gravity. Phys. Rep. 119 (1985) 1.

\bibitem{birdav} N.D. Birrell and P.C.W. Davies,
{\sl Quantum Fields in Curved Space}
(Cambridge Univ. Press, Cambridge, 1982).

\bibitem{betor}
A.S. Belyaev, I.L. Shapiro,
Phys.Lett.  B425 (1998) 246;
Nucl.Phys. B543 (1999) 20.

\bibitem{guhesh}    G. de Berredo-Peixoto,
J.A. Helayel-Neto and  I. L. Shapiro,
JHEP 02 (2000) 003.

\bibitem{torsi}
I.L. Shapiro,
Phys. Repts. 357 (2002) 113.

\bibitem{chris} S.M. Christensen,
J. Phys. A: Math. Gen. (1980). 13 3001.

\bibitem{Kost-post-New} Q. G. Bailey and V. A. Kosteleck\'y,
Signals for Lorentz Violation in Post-Newtonian Gravity,
[gr-qc/0603030].


\end{thebibliography}
\end{document}